\documentclass[doublecol]{epl2} 

\usepackage{graphicx}
\usepackage{amsmath}				
\usepackage{amssymb}				
\usepackage{amsfonts}				
\usepackage{amstext}

\newcommand{\bra}[1]{\langle #1|}
\newcommand{\ket}[1]{|#1\rangle}
\newcommand{\braket}[2]{\langle #1|#2\rangle}
\newcommand{\vac}{|\mbox{vac}\rangle}
\newcommand{\ev}[1]{\langle #1 \rangle}

\title{Stochastic mean-field theory for the disordered Bose-Hubbard model}
\author{Ulf Bissbort\inst{1} \and Walter Hofstetter\inst{1}}
\shortauthor{U. Bissbort and W. Hofstetter}

\institute{                    
  \inst{1} Institut f\"ur Theoretische Physik, Johann Wolfgang Goethe-Universit\"at - 60438 Frankfurt/Main, Germany
}

\abstract{
We investigate the effect of diagonal disorder on bosons in an optical lattice described by an Anderson-Hubbard model at zero temperature. It is known that within Gutzwiller mean-field theory spatially resolved calculations suffer particularly from finite system sizes in the disordered case, while arithmetic averaging of the order parameter cannot describe the Bose glass phase for finite hopping $J>0$. Here we present and apply a new \emph{stochastic} mean-field theory which captures localization due to disorder, includes non-trivial dimensional effects beyond the mean-field scaling level and is applicable in the thermodynamic limit. In contrast to fermionic systems, we find the existence of a critical hopping strength, above which the system remains superfluid for arbitrarily strong disorder.}

\begin{document}
\maketitle

Ever since the seminal paper by Fisher {\it et al.} \cite{Fisher_Fisher}, the disordered Bose Hubbard model has been the subject of theoretical and experimental investigation. 
In particular, the realization of the superfluid-Mott insulator transition in a gas of ultracold bosonic atoms in an optical lattice \cite{Bloch_Greiner_Mott_transition} has sparked a new wave of research on this field. 
In contrast to typical solid state systems, optical lattices allow the introduction of various types of disorder in a highly controlled manner \cite{Lewenstein_Disorder_review}. Experimentally, several techniques have been implemented, such as speckle laser patterns  \cite{Inguscio_speckle,Aspect_speckle, aspect_1d_disorder,DeMarco_3d_disorder}, multichromatic lattices with non-commensurate wavelengths \cite{Roth_Inguscio_two_color_lattice, Inguscio_non_interacting} or an additional species of atoms tunneling at a considerably lower rate \cite{Castin_binary_disorder,Sengstock_Esslinger_two_species}. In a recent experiment \cite{DeMarco_3d_disorder} it was possible to generate the first 3D fine-grained disorder potential using a speckle laser, providing an excellent realization of the 3D disordered Bose-Hubbard model.
Various theoretical methods have previously been employed to investigate the transitions between 
Mott insulator (MI), Bose glass (BG) and superfluid (SF), such as Quantum Monte Carlo \cite{Krauth_QMC,Rieger_QMC,Haas_QMC,Svistunov_QMC,Zimanyi_QMC}, exact diagonalization \cite{Roth_Inguscio_two_color_lattice,Sengstock_exact_diag,Davidson_exact_diag,Tsubota_exact_diag},  renormalization group \cite{Singh_RG}, density matrix renormalization group \cite{Zwerger_DMRG} and mean-field approximations \cite{Rokhsar_Kotliar,Krauth_Gutzwiller,Fisher_Fisher,Buonsante_1d_MFT,Buonsante_2d_do_trap,Lewenstein_Bose_anderson_glass,Ramakrishnan_MFT_RPA,Sheshadri,Graham_arithmetic_MFT,Graham_self_consistent_MFT}. However,  spatially resolved calculations on disordered lattices suffer from finite size effects in the vicinity of phase borders, where the physics is dominated by rare events, while an arithmetically averaged mean-field theory is incapable of describing the Bose glass phase at any finite hopping amplitude and $T=0$ \cite{Graham_arithmetic_MFT}. 

Ultracold bosonic atoms in a sufficiently deep optical lattice at moderate filling are well described by the single band Bose-Hubbard (BH) Hamiltonian \cite{Jaksch_cold_atoms}
\begin{equation}
\mathcal{H}_{\mbox{\tiny BH}}=-J \sum_{\ev{i,j}} ( b_i^\dag b_j^{\phantom{\dag}} +  b_j^\dag b_i^{\phantom{\dag}})  + \sum_i (\epsilon_i-\mu){b}_i^\dag b_i^{\phantom{\dag}} + \frac{U}{2}\sum_{i} b_i^\dag b_i^\dag b_i^{\phantom{\dag}} \, b_i^{\phantom{\dag}}
\end{equation}

\begin{figure}
\onefigure[width=0.8\linewidth]{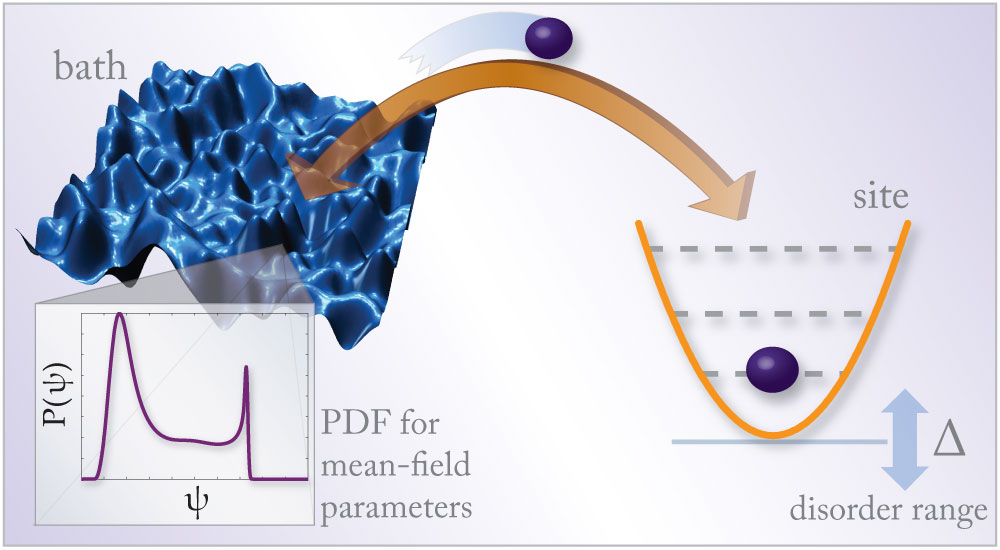}
\caption{Within SMFT, the multiple site lattice model is approximated by  an effective single-site problem, where a site is coupled to a bath of mean-field parameters (MFPs). Disorder-induced fluctuations of the MFPs are accounted for by a statistical distribution $P(\psi)$.}
\label{fig:SMFT_cartoon}
\end{figure}

\noindent 
where $b_i^\dag$ is the bosonic creation operator for an atom in the lowest Wannier state at lattice site $i$, $J$ is the hopping amplitude, $U$ describes a short-ranged interaction, $\ev{i,j}$ denotes the sum over all pairs of neighboring sites and $\mu$ is the chemical potential. In performing the lowest band approximation, it is implicitly assumed that the temperature is sufficiently low to suppress all contributions of states in higher bands. Since all system properties only depend on the ratios of the energy scales, we choose to work in units of $U=1$. Diagonal disorder is parametrized by on-site energies $\epsilon_i$, which we will assume to be independently and identically distributed \footnote{This is a justifiable approximation considering new experimental techniques \cite{DeMarco_3d_disorder}} according to a box distribution $p(\epsilon)=\Theta(\Delta/2 - |\epsilon|)/\Delta$ with the disorder strength parameter $\Delta$, while the hopping amplitude $J$ is assumed to be site-independent. On a mean-field level, the superfluid-insulator transition can be captured by the variational bosonic Gutzwiller ansatz \cite{Rokhsar_Kotliar}
\begin{equation}
	\ket{\mbox{GW}}=\prod_i \left[ \sum_{n=0}^\infty \frac{f_n^{(i)}}{\sqrt{n!}} (b_i^\dag)^n \right] \; \vac ,
\end{equation}
which is a good approximation in high dimensions (coordination number $Z$) and delivers the exact ground state in the weak tunneling limit $J\to 0$, as well as in the non-interacting limit $U\to 0$ \cite{Zwerger_MF}.
The mean-field ground state is determined by minimizing the energy expectation value $\bra{\mbox{GW}} \mathcal{H}_{\mbox{\tiny BH}} \ket{\mbox{GW}}$ with respect to all the amplitudes $\{f_n^{(i)}\}$ under the constraint $\braket{\mbox{GW}}{\mbox{GW}}=1 $. For a pure system ($\Delta=0$)  at $T=0$ this leads to the same ground state as that of the site decoupled mean-field Hamiltonian \cite{Ramakrishnan_MFT_RPA} 
\begin{equation}
\begin{split}
	\mathcal{H}_{\mbox{\tiny MF}}=\sum_i \Bigl[ -J \sum_{\mbox{\scriptsize n.n.}j}(\psi_j^* b_i^{\phantom{\dag}} + \psi_j^{\phantom{\dag}} b_i^\dag -\psi_j^*\psi_i^{\phantom{*}} )  \\
	   +(\epsilon_i-\mu) {b}_i^\dag b_i^{\phantom{\dag}} + \frac{U}{2}  b_i^\dag b_i^\dag b_i^{\phantom{\dag}} \, b_i^{\phantom{\dag}} \Bigr],
	   \end{split}
\end{equation}
where the mean-field parameters (MFPs) $\psi_i=\ev {b_i}$ are determined self-consistently. 
In the ground state, all MFPs have the same complex phase and, due to the global $U(1)$-symmetry of $\mathcal{H}_{\mbox{\tiny BH}}$, can be chosen to be real and non-negative. 

For $\Delta=0$ a vanishing order parameter (defined as any of the identical MFPs) indicates an insulating state of the system, while a finite value indicates a Bose condensed (SF) phase. The spatially resolved, self-consistent Gutzwiller approach has been used to study disordered bosons in optical lattices within finite size simulations  \cite{Buonsante_1d_MFT,Buonsante_2d_do_trap}, but suffers of the disadvantage that it overestimates phase coherence, especially in the SF phase in the vicinity of the phase borders and cannot describe the BG phase at $T=0$. However, simulating disorder effects in finite systems is a delicate problem, since rare events may strongly influence physical observables such as the excitation spectrum. 

Here we present and apply a \emph{stochastic} mean-field theory (SMFT) for disordered bosons, which extends the self-consistent Gutzwiller approach to the thermodynamic limit and is free of finite-size effects. Its numerical efficiency and absence of finite size effects make this method a good candidate for future analysis of more complex systems, such as multicomponent gases in disordered lattices. This is achieved by a probabilistic description of an infinite system, by using a probability density function (PDF) $P(\psi)$ to allow disorder-induced fluctuations of the MFPs. Probabilistic descriptions have been successfully applied to disordered systems previously, such as by \cite{Vojta_MFT} for antiferromagnets. 

\begin{figure}[t]
\onefigure[width=0.97\linewidth]{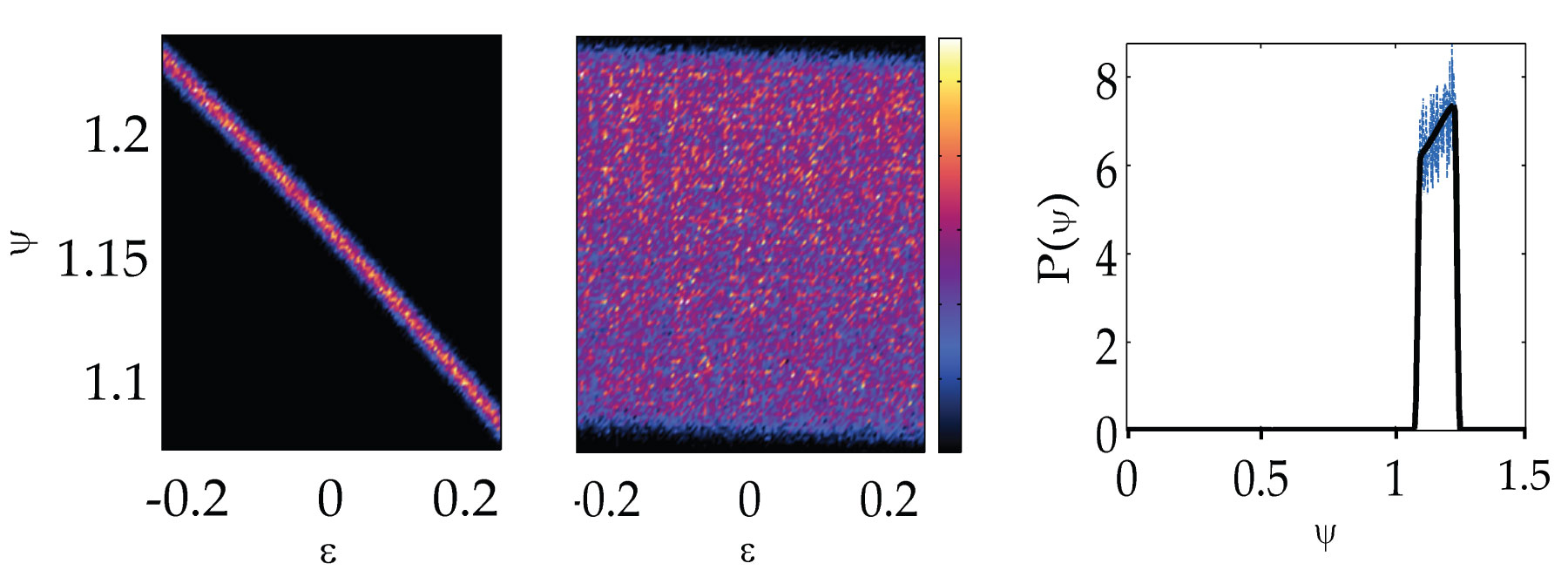}
\caption{Results from a spatially resolved, bosonic Gutzwiller calculation (27 000 sites) in the deep SF regime at $J=0.067$, $Z=6$, $\mu=1$, $\Delta=0.5$, where this method becomes exact (Gross-Pitaevskii regime with depletion). The left figure is a color-coded histogram (corresponding to the 2D probability density function), showing the strong correlation between the on-site energy $\epsilon$ and the on-site MFP $\psi$ (correlation coefficient $-0.9964$), which is not neglected within SMFT. The central figure is a color-coded histogram of the on-site energy and nearest neighbor MFPs, which are only weakly correlated (correlation coefficient $-0.0349$) and neglected within SMFT. The right figure is a comparison of a self-consistently determined distribution from SMFT for these parameters, compared to a 1D histogram of MFPs from the spatially resolved Gutzwiller calculation (fluctuations originate from the finite system size). Excellent agreement is achieved in this limit.}
\label{fig:comp_SRMF}
\end{figure}
The many-particle, multiple-site Bose-Hubbard model is thereby reduced to an effective single-site mean-field problem, the solution of which entails the self-consistent determination of $P(\psi)$. To derive the SMFT self-consistency condition for $P(\psi)$, we pursue the self-consistent mean-field approach and assume that the on-site energy of a given site and the MFPs of the neighboring sites are uncorrelated. To justify this approximation, we performed a spatially resolved Gutzwiller calculation, which is known to give a good approximation for the ground state for weak interactions $U\ll JZ$. In Fig.\ref{fig:comp_SRMF} we plot a histogram, showing that the correlation between an on-site energy and the nearest neighbor MFP is weak and the above approximation is justified for weak interactions.

Considering an arbitrary site $i$ with energy $\epsilon_i$, the further quantity determining its mean-field ground state is the scaled sum of MFPs from the neighboring sites 
$\eta= J\sum_{\mbox{\scriptsize n.n.}j=1}^Z \psi_j$
distributed according to the PDF
 \begin{equation}
Q(\eta)=\int_0^\infty d\psi_1 \, P(\psi_1) \ldots \int_0^\infty d\psi_Z \, P(\psi_Z) \; \delta \Bigl( \eta- J\sum_{m=1}^Z \psi_m \Bigr)
\end{equation} 
Hence, once the self-consistent solution $P(\psi)$ is known, any disorder averaged expectation value of a single-site operator can be expressed as 
\begin{equation}
	\overline {\ev{\hat{A}}} = \int d\epsilon \, p(\epsilon) \int d\eta \, Q(\eta)\, \bra{\mbox{gs}(\epsilon,\eta)} \hat A \ket{\mbox{gs}(\epsilon,\eta)},
\end{equation}
where $\ket{\mbox{gs}(\epsilon,\eta)}$ is the ground state of 
\begin{equation}
\label{mf_hamiltonian}
\mathcal{H}=\eta(b^\dag+b)+(\epsilon-\mu) {b}^\dag b + \frac{1}{2}  b^\dag b^\dag b \, b.
\end{equation}

The self-consistency condition requires that if the on-site energy $\epsilon$ is randomly chosen from $p(\epsilon)$ and $Z$ MFPs are drawn from $P(\psi)$ to account for an effective tunneling from the nearest neighbors (or equivalently $\eta$ is drawn from $Q(\eta)$), the calculated expectation values $\ev b$ have to be distributed according to the initially assumed PDF $P(\psi)$ (illustrated in Fig.\ref{fig:SMFT_cartoon}). This can be expressed by the self-consistency equation
\begin{align}
\begin{split}
\label{sc_eqn}
	\int_0^\infty d\eta \, Q(\eta) \, \tilde{P}_\eta(\psi)=P(\psi), \hspace{15mm} \\
\tilde P_\eta(\psi):=\frac{d}{d\psi} \int d\epsilon \: p(\epsilon) \: \Theta\left( \psi - \bra{\mbox{gs}(\epsilon,\eta)} b \ket{\mbox{gs}(\epsilon,\eta)} \right)
\end{split}
\end{align} 
is the conditional probability density for a site having the MFP $\psi$ if the external coupling $\eta$ is given and the disorder energy is distributed according to $p(\epsilon)$.

\begin{figure}[t]
\onefigure[width=0.97\linewidth]{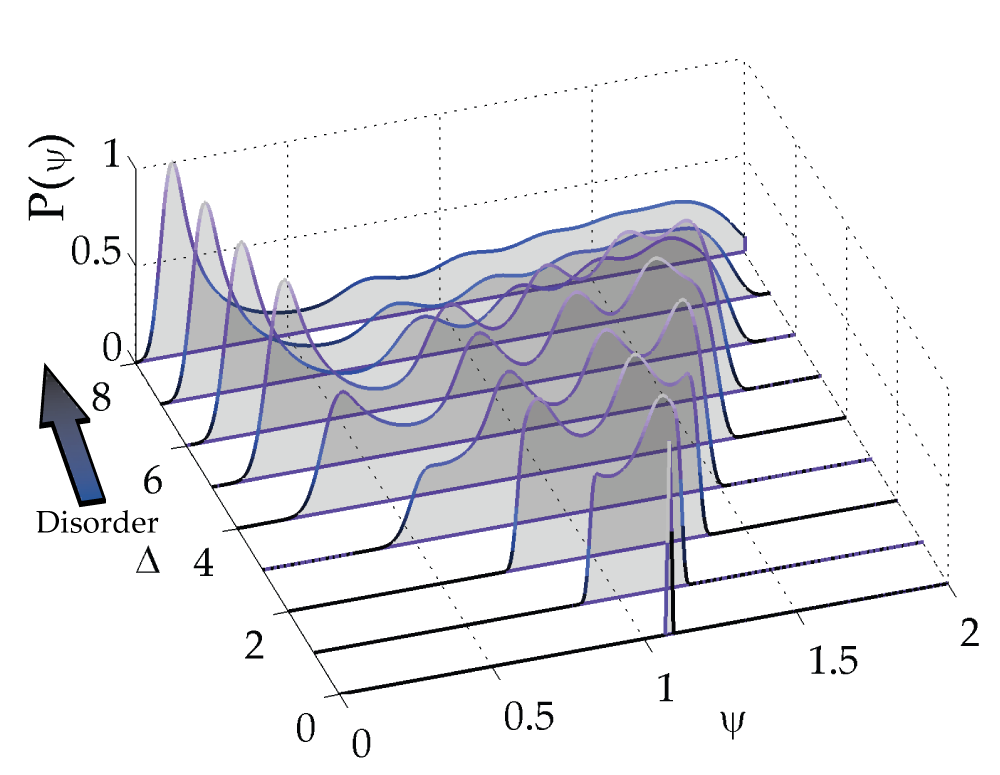}
\caption{Self-consistently determined MFP probability density distributions at fixed  $\mu=1.0$, $J = 0.05$, $Z = 6$  and increasing disorder strength $\Delta$. The distributions are normalized to their maximum value for visual clarity. For $\Delta \to 0$ the PDF converges to a shifted $\delta$-distribution, recovering to the well-known usual MFT for the pure system. With increasing $\Delta$ the disorder induces fluctuations in the MFPs, i.e. $P(\psi)$ broadens.}
\label{fig:dists_delta}
\end{figure}

As a next step it is of interest to characterize the different phases once the distribution $P(\psi)$ has been determined. The condensate fraction within the SMFT is given by
\begin{equation}
f_c={\overline{\ev{b}^2}}/{\overline{\ev{b^\dag b}}},
\end{equation} 
which reveals that the system is in the superfluid phase as soon as $P(\psi)\neq \delta(\psi)$ (i.e. $f_c>0$). Rigorously speaking, this should be referred to as the condensed phase, as a finite value of the average MFP indicates the existence of a macroscopically occupied single-particle state. To distinguish the MI from the BG phase, a further quantity, such as the compressibility $\kappa=\frac{\partial \overline{ \langle n \rangle}}{\partial \mu }$ (which vanishes only in the MI), has to be considered.

\begin{figure*}[]
\begin{center}
\includegraphics[width=0.94\linewidth]{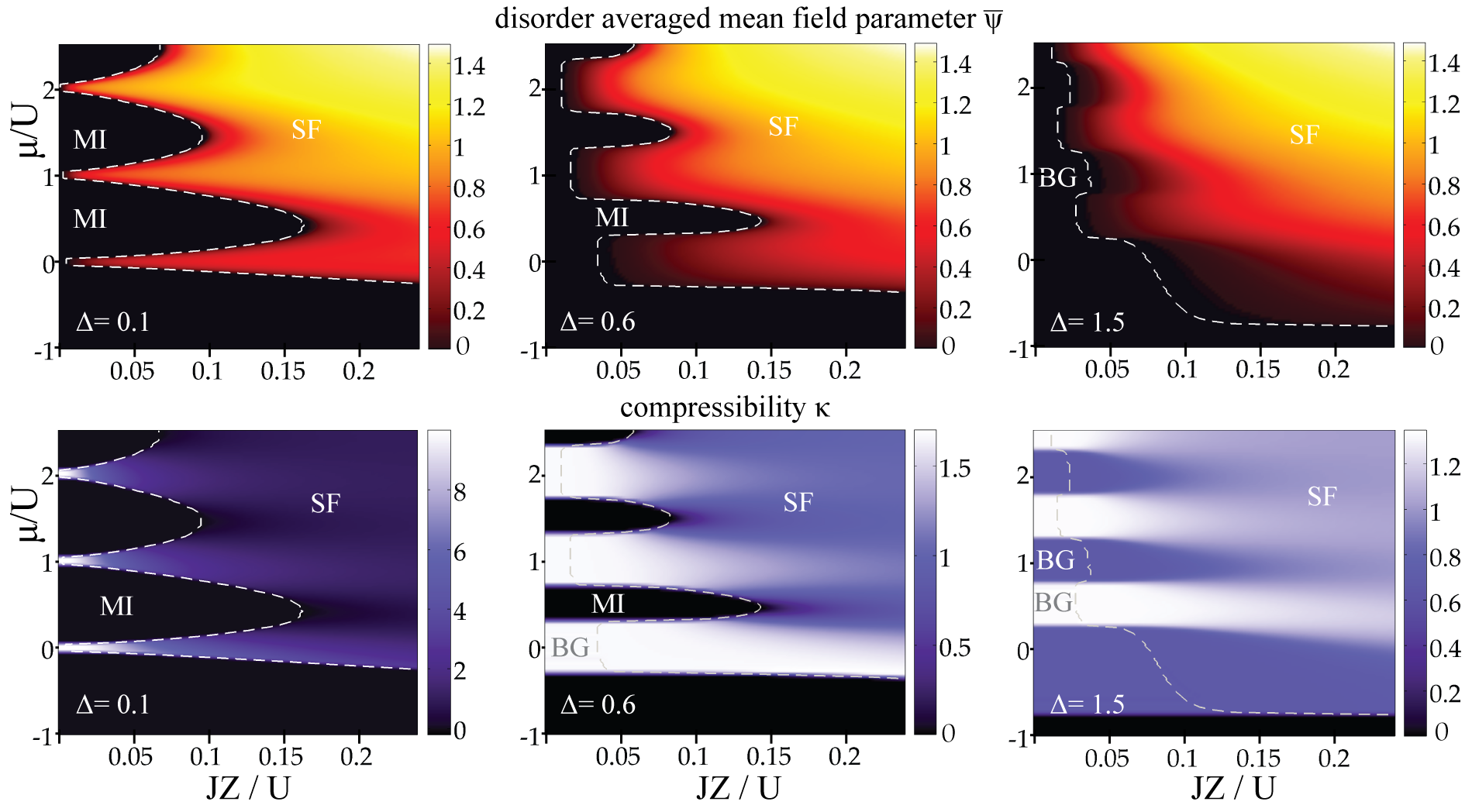}
\vspace{-0.3cm}
    \caption
    {\label{fig:Phase_diags}
    Plots of the arithmetic mean order parameter $\overline \psi=\int d\psi \, P(\psi) \, \psi$ (upper row) and the compressibility $\kappa$ (lower row) show the stochastic mean-field phase diagram for three different disorder strengths and $Z=6$. The white lines indicate the phase borders and black corresponds to the value zero for all plots. For weak disorder ($\Delta=0.1$) the phase diagram closely resembles that of a pure system. With increasing disorder the Mott insulating regions ($\overline \psi=0, \, \kappa=0$) shrink and are completely replaced by the Bose glass ($\overline \psi=0, \, \kappa>0$) and superfluid ($\overline \psi>0, \, \kappa>0$) phases at $\Delta=1$. The areas for $\mu<0$ with $\overline \psi=0$ and $\kappa=0$ correspond to the vacuum.  (color online)
\vspace{-4mm}
}
\end{center}
\end{figure*}

For general parameters $\mu,\, J,\, Z,\,$ and $\Delta$, the distribution $P(\psi)$ fulfilling the self-consistency condition (\ref{sc_eqn}) cannot be determined analytically. We determine  $P(\psi)$ by a numerical iterative procedure, beginning with any distribution, other than the insulating solution $P_{\mbox{\tiny MI}}(\psi)= \delta(\psi)$, which always fulfills the self-consistency equation as it is a fixed point of the iterative mapping. We have verified numerically that there always exists a unique attractive self-consistent solution, which we identify as the physical distribution (which also minimizes the grand canonical potential). A multiple grid discretization procedure, yielding a high resolution to detect the superfluid-insulator transition at small values of $\psi$ was used for the numerical tabulation of $P(\psi)$.

Typical results for such distributions are shown in Fig.\ref{fig:dists_delta} for a variety of increasing disorder strength values $\Delta$. At $\Delta=0$ the distribution $P(\psi)$ is a $\delta$-function at the value of $\psi$ corresponding to the solution of the usual Gutzwiller bosonic MFT (in the disorder-free case). In the presence of disorder ($\Delta>0$), $P(\psi)$ acquires a finite width in the SF phase. This can be understood to have two origins: Fluctuations of the on-site energy $\epsilon$ necessarily lead to a variation in the calculated MFP $\ev b$ (for non-zero MFPs from the neighboring sites).  Subsequent additional fluctuations in the MFPs furthermore enhance the fluctuations of $\ev b$. By decreasing the hopping strength for fixed disorder, we find that the system is always driven into an insulating state with $\overline \psi=0$. Care has to be taken at the SF/BG transition, where it has to be ensured that the distribution is independent of the numerical discretization.
Since no averaging of the MFPs is performed, the results of the SMFT depend nontrivially (beyond $JZ$-scaling) on the dimensionality of the system: with increasing dimensionality the BG region in the phase diagram gives way to the superfluid phase. Using the properties of the convolution, it can be seen that in the limit of infinite dimensions the arithmetically averaged MFT is recovered, where the BG can only exist at $J=0$. In this sense, the existence of the BG depends on fluctuations of the MFPs, which are accounted for by the probabilistic description within the SMFT. Since the relative fluctuations of the MFPs decrease with the dimension of the system, the BG phase exists in a larger region of the phase diagram (scaled with $JZ$) in low dimensions. 
Furthermore, the question of a critical dimension for the crossover from a direct transition between the MI and SF and an indirect transition, always occurring via the BG phase, arises\footnote{After completing this work, a preprint also addressing this question appeared \cite{Pollet_prokojev_svistunov_troyer}}. The latter scenario is well-established in one and two dimensions, while in the limit of high dimensions, where SMFT is known to become exact, a direct transition at the tip of the Mott lobes is predicted.

\begin{figure}
\onefigure[width=0.9\linewidth]{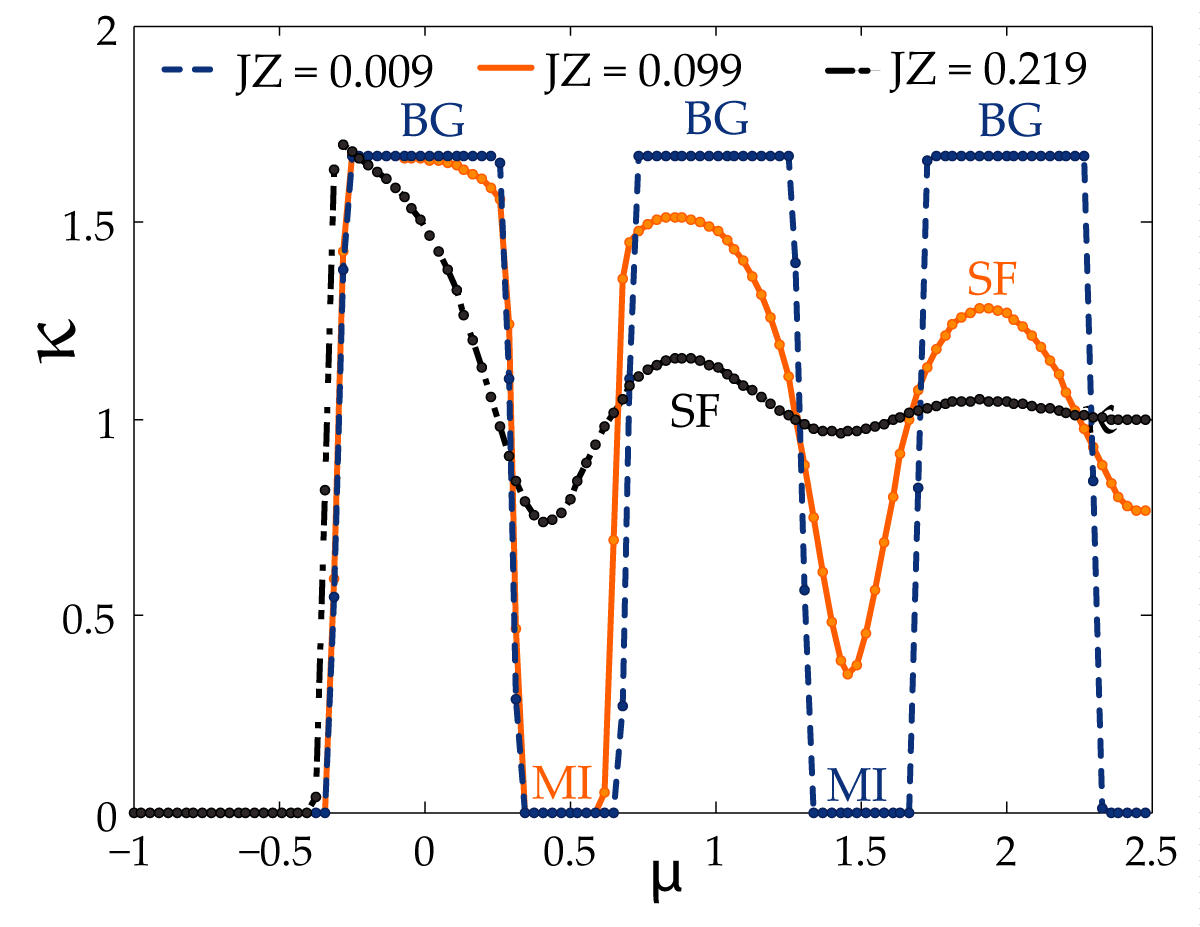}
\caption{The compressibility $\kappa=\frac{\partial n}{\partial \mu}$ for intermediate disorder strength $\Delta=0.6$ and $Z=6$ as a function of $\mu$. For small hopping parameters (blue line $JZ=0.009$) the system is driven through an alternating sequence of MI and BG phases. For stronger hopping (red line $JZ=0.099$) the system enters and remains in the superfluid phase from some value of $\mu$ onwards, where it varies smoothly (color online).}
\label{fig:kappa_along_mu}
\end{figure}
The resulting phase diagrams for a 3D lattice are shown in Fig.\ref{fig:Phase_diags}.
The transitions between the MI, BG and SF phases are of second order, since the density $\overline{\ev{n(\mu)}}$ varies continuously, but the compressibility is discontinuous at the transition points, as shown in Fig.\ref{fig:kappa_along_mu}. In the BG phase, the compressibility is proportional to the density of sites at positive integer values of the effective chemical potential $\mu'=\mu-\epsilon$. Therefore it takes on constant positive values on finite intervals of $\mu$ in the BG phase for a box disorder distribution, while it varies continuously within the SF phase, as can be seen in Fig.\ref{fig:kappa_along_mu}.

\begin{figure}
\onefigure[width=0.9\linewidth]{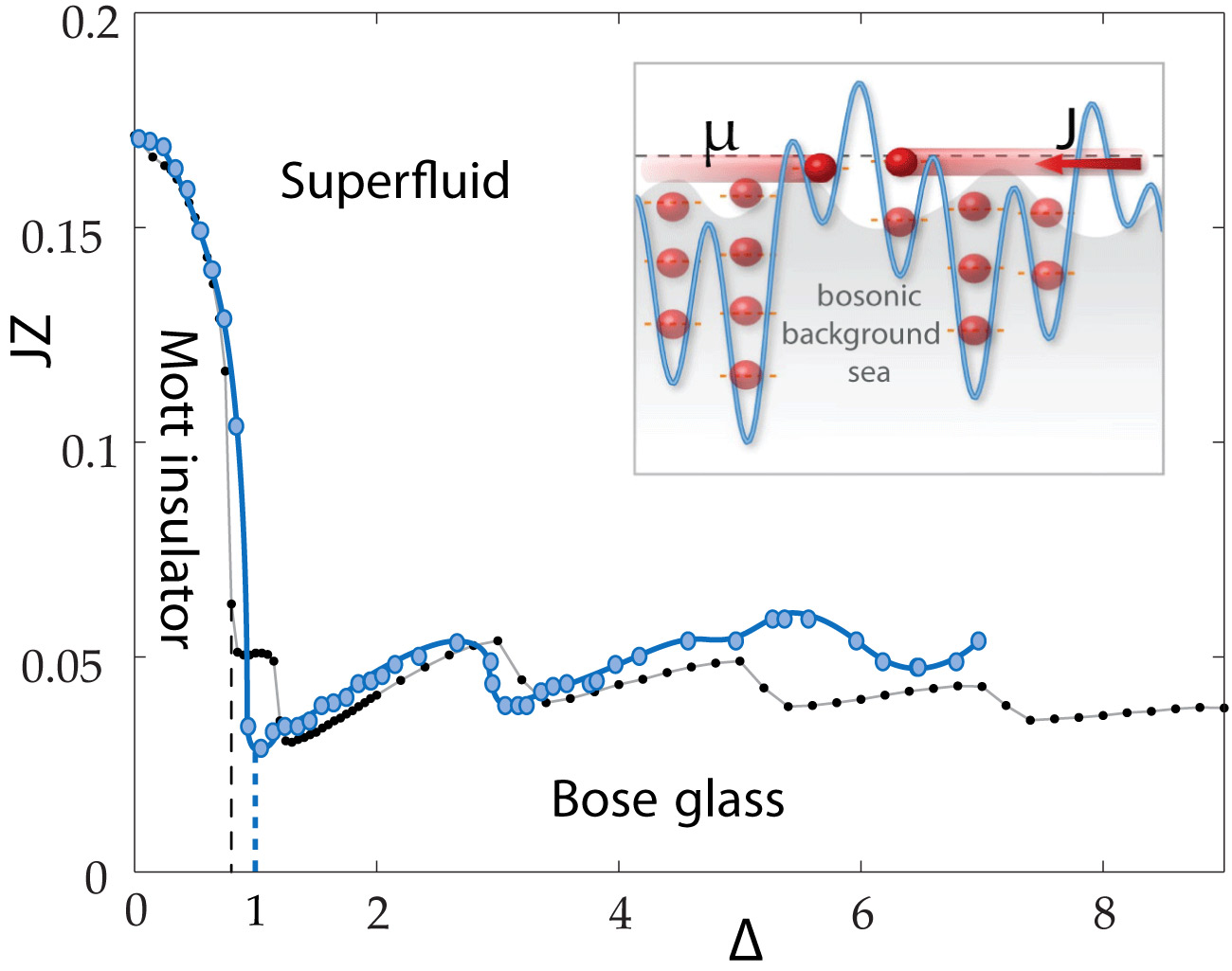}
\caption{SMFT phase diagram in the  $JZ-\Delta$-plane for fixed $\mu=0.4$ (gray line, small black dots) and fixed density $n=1$ (blue line and larger circles, with error $\pm 0.003$ in $JZ$). Inset: illustration of the bosonic background screening the strongly disordered spatial potential. The MI/BG phase borders occur at constant $\Delta$, since the mean-field state cannot depend on $JZ$ if all MFPs $\psi$ vanish. (color online)}
\label{fig:JZ_Delta_PD}
\end{figure}

The SMFT predicts that for every fixed value of $\mu$, there exists a certain value $JZ_{c}(\mu)$ above which the system is always in a superfluid state, independent of the disorder strength $\Delta$. Typical phase diagrams in the $JZ$-$\Delta$-plane for constant $\mu=0.4$ and for integer filling $n=1$ are shown in Fig.\ref{fig:JZ_Delta_PD}. With increasing disorder strength at constant $\mu$, we find that the superfluid-insulating phase border moves to smaller values of $JZ$, while fluctuating in $\Delta$ with a periodicity of $2U$. This periodic behavior at small $\mu/U$ originates from the relative weight of Mott insulating lobes within the disorder interval.

This can be understood qualitatively by interpreting the disordered BH model in terms of the pure phase diagram, where the disorder corresponds to a whole range of different values of $\mu$ contributing to each point in the disordered phase diagram. Deep wells (i.e. lattice sites with low effective chemical potential) are successively filled up with a suitable number of particles, forming a bosonic sea which effectively screens the disordered potential - on top of which it is energetically favorable for the remaining particles to delocalize (illustrated in the inset in Fig.\ref{fig:JZ_Delta_PD}). In the limit of strong disorder, it should be kept in mind that these predictions are only valid in a regime where the single band Bose-Hubbard model is justified. 
 
In conclusion, we have developed a stochastic mean-field approach to the disordered Bose-Hubbard model in three spatial dimensions. By working with the full distribution function of MFPs without averaging these, we are able to describe the Bose glass phase and the underlying localization of bosons within a \emph{local} approach at $T=0$, which becomes rigorous in the limit of high spatial dimensions. In contrast to spatially resolved Gutzwiller calculations, which predict condensation into a superposition of distant localized single-particle states (i.e. overestimate the coherence) in the BG regime, the SMFT does not suffer from this problem. Furthermore, we observe a direct transition between Mott insulator and superfluid in the presence of disorder and find that superfluidity persists above a critical hopping strength for arbitrarily strong disorder at fixed $\mu$, due to screening of strong potential fluctuations by accumulation of bosons. These findings and quantitative predictions are of immediate relevance for current \cite{DeMarco_3d_disorder} and upcoming experiments on disordered bosons in optical lattices.  

Being a single-site theory, the SMFT is numerically less demanding than other methods describing the BG transition, such as Quantum Monte Carlo calculations or renormalization group studies. This allows an extension of the method to more complex systems, such as multi-component mixtures. However, the single-site nature has the drawback that spatial information is lost. The stochastic approach chosen here may furthermore be extended in future to describe fluctuations of different origin, such as thermal or quantum fluctuations.

\acknowledgments
We thank I. Bloch, B. DeMarco, E. Demler, A. Pelster, M. Snoek, R. Thomale and W. Zwerger for useful discussions. 
This work was supported by the Deutsche Forschungsgemeinschaft via Forschergruppe FOR 801.
Support by the \textit{Studienstiftung des deutschen Volkes} is gratefully acknowledged by UB.

\end{document}